\documentclass[a4paper]{article}
\usepackage{graphicx}
\usepackage{subfig}

\title{Octonions}
\author{Jonathan Hackett\thanks{Email address:
jhackett@perimeterinstitute.ca}\\
Perimeter Institute for Theoretical Physics,\\
31 Caroline St. N., Waterloo, Ontario N2L 2Y5, Canada, and \\
Department of Physics, University of Waterloo,\\
Waterloo, Ontario N2J 2W9, Canada\\
 \and Louis H. Kauffman\thanks{Email address:
kauffman@uic.edu}\\
Math, UIC\\
851 South Morgan Street\\
Chicago, IL 60607-7045, United States\\}

\begin{document}

\maketitle

\begin{abstract}
In this paper we review the topological model for the quaternions based upon the Dirac string trick. We then extend this model, to create a model for the octonions - the non-associative generalization of the quaternions.
\end{abstract}

\section{Introduction}
In this paper we give a topological/combinatorial model of the octonions that is an extension of
an already-existing model of the quaternions. The model of the quaternions \cite{KP} is based on the
belt trick or Dirac string trick, and will be reviewed in the first section of this paper. We then proceed
to find a corresponding model for the octonions. The octonions are a non-associative generalization of the quaternions that have been used directly and speculatively in physics for some time
\cite{Furey:2010fm,catto,gursey,baez,okubo}. We hope that the present model will lead to new
physical insight.
\bigbreak

Recall first the definition of the the quaternions. The quaternions are an associative algebra over the real numbers generated by linearly independent elements
$$1, i, j, k$$
with $$i^2 = j^2 = k^2 = ijk = -1.$$ From these equations, it follows that $ij= k, jk=i, ki = j$
and $ji = -k, kj = -i, ik = -j.$  In general, a quaternion $A$ has the form
$$A = a + bi + cj + dk$$ where $a,b,c,d$ are real numbers. The {\it conjugate} $\bar{A}$ of $A$ is defined by the equation $$\bar{A} = a - bi - cj - dk$$ and has the property that
$$A \bar{A} = a^2 + b^2 + c^2 + d^2,$$ showing that non-zero quaternions have multiplicative inverses.
\bigbreak

The octonions are a non-associative algebra obtained by adding a new element $L$ to the quaternions.
If $A,B,C,D$ are quaternions, then the products in the octonions are all determined by the following
formula $$(A + LB)(C + LD) = (AC - D\bar{B}) + L(CB + \bar{A}D).$$
Another way of putting this, that is useful for our purposes is the following:
Suppose that $$x,y \in \{ i,j,k\}.$$ Then
$$LL = -1,$$
$$xL = -Lx,$$
$$(Lx)y = L(yx),$$
$$(Lx)(Ly) = yx,$$
$$x(Ly) = -L(xy).$$
We will use this form of the identities for octonion multiplication to check the properties of our
model of the octonions.
\bigbreak

\section{The Quaternions}

Through a topological property, commonly referred to as the {\it Dirac string trick},  we can construct a physical/topological  model of the quaternion group. In this model one takes a geometric object in
Euclidean three-dimensional space and attaches a belt (i.e. a space homeomorphic to the cross product of a unit interval with itself) to the object and to a reference point. The reference point is often taken to be the north pole of a two dimensional sphere surrounding the object.  Rotations of the object carry the belt along, twisting it without causing self-intersections or singularities. A $2 \pi$ rotation of the object about
an axis causes a twist to appear in the belt that cannot be removed by topological isotopy of the belt,
leaving the endpoints of the belt fixed to the object and to the reference point. But a $4 \pi$ rotation
gives a state of the belt that can be isotoped to its original (untwisted) state by isotopy fixing the endpoints. This topological fact is usually called the belt trick or Dirac string trick.
\bigbreak

 If we attach a belt to an object $O$ with symmetry group $G$ in $SO(3)$ ((the group of orientation  preserving rotations of Euclidean three-space), then each symmetry of the object acquires two possible states: The two states differ by a $2 \pi$ twist of the belt. The result is a
doubling of the symmetry group of the object to a new group $\hat{G}.$  If we take the object to be a {\it belt buckle} (i.e. a rectangle), then the symmetry group is $G = Z/2Z \times Z/2Z$ of order $4$, and  $\hat{G}$ is the eight-element quaternion group.
\bigbreak

Previously, this property has been used to construct a model for the quaternions \cite{KP} by working specifically with belt and belt-buckle.  The reason behind the working of such models has to do with the fact that one has a double covering $p: SU(2) \longrightarrow SO(3)$ where $SU(2)$ denotes the
unitary $2 \times 2$ matrices of determinant one. $SU(2)$ is isomorphic with the quaternions of unit length, and one can show that the group $\hat{G}$ is isomorphic with the inverse image of $G$
under this double covering maping. Thus $\hat{G} = p^{-1}(G).$ The topology of the belt returning to
its identity state from a $4 \pi$ rotation is a consequence of the fact that the fundamental group of the Lie group $SO(3)$ is of order two. The relationship with $SU(2)$ demonstrates clearly the relation between the quaternions and the physical situations in which they arise in naturally. In particular, the non-triviality
of the $2 \pi$ rotation and the triviality of the $4 \pi$ rotation corresponds to the fact in quantum mechanics that the wave function of a fermion changes by a sign if the system undergoes a rotational
symmetry of $2 \pi.$ Symmetries of the system are mapped to unitary transformations in quantum
mechanics, and so corresponnd to lifts to the double cover of $SO(3).$
\bigbreak

To construct this model of the quaternions using belt and buckle,  we consider a belt that has been fixed to a wall with the non-buckle end. We consider $\pi$ rotations of the belt buckle about the three standard cartesian axes which we correspond to the three quaternionic roots of $-1$: $i$,$j$, and $k$.  We then get that carrying out $j$ after $i$ yields the same result as performing $k$ - likewise for any other combination of $i$,$j$ and $k$ - with $-x$ equivalent to $x$ but with the twisting of the belt in the opposite direction.  We also get that carrying out any operation twice yields a belt that is twisted around by a full $2\pi$, which we then call $-1$.  The final step in getting our model is to recall the Dirac belt trick which tells us that if we perform $-1$ twice - giving us a $4\pi$ rotation - we can remove all of the twisting without rotating the belt buckle.  We use this operation to remove any extraneous twisting, and find that our operations exactly correspond to all of the elements of the quaternions.  We note that the operations are performed from left to right along a string of elements.

\begin{figure}[!h]
\centering
\subfloat[]{\includegraphics[height=1.8in]{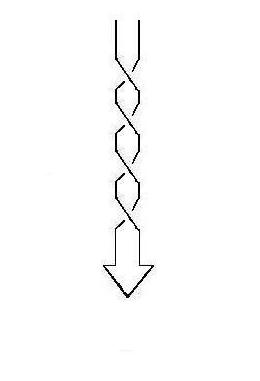}}\ \
\subfloat[]{\includegraphics[height=1.8in]{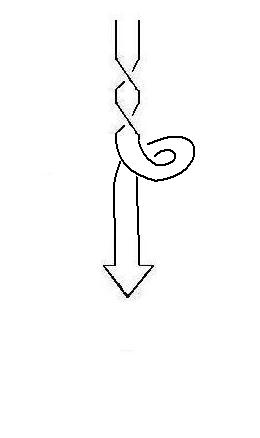}}\ \
\subfloat[]{\includegraphics[height=1.8in]{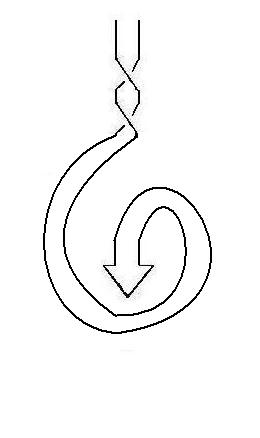}}\ \
\subfloat[]{\includegraphics[height=1.8in]{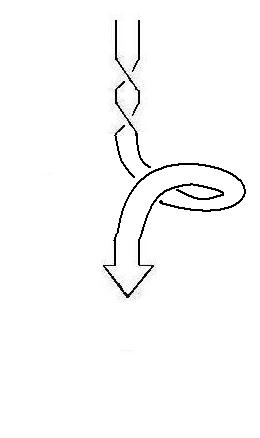}}\ \
\subfloat[]{\includegraphics[height=1.8in]{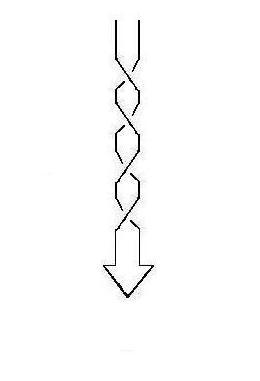}}\ \
\subfloat[]{\includegraphics[height=1.8in]{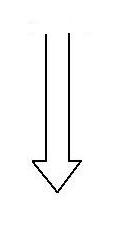}}
\caption{Step-by-step demonstration of the Dirac string trick on a belt}
\end{figure}

\begin{figure}[!h]
\centering
\subfloat[$1$]{\includegraphics[height=1.6in]{qident.jpg}}\qquad
\subfloat[$i$]{\includegraphics[height=1.6in]{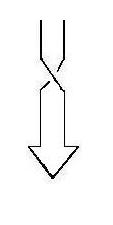}}\qquad
\subfloat[$j$]{\includegraphics[height=1.6in]{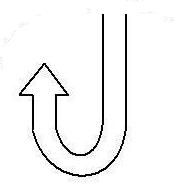}}\qquad
\subfloat[$k$]{\includegraphics[height=1.6in]{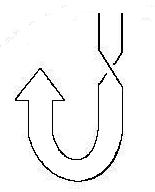}}
\caption{Belt states of the identity and the quaternions $i$, $j$ and $k$} \label{quaternions}
\end{figure}

We can observe that beginning with $i$ and performing $j$ we reach $k$.
\begin{figure}[!h]
\centering
\subfloat[]{\includegraphics[height=1.6in]{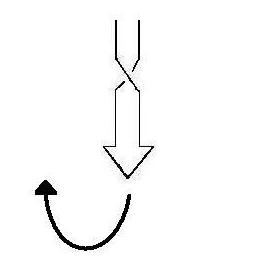}}\qquad
\subfloat[]{\includegraphics[height=1.6in]{qk.jpg}}
\caption{$ij=k$}
\end{figure}
And that beginning with $j$ and performing $i$ we instead reach $-k$.
\begin{figure}[!h]
\centering
\subfloat[]{\includegraphics[height=1.6in]{qj.jpg}}\qquad
\subfloat[]{\includegraphics[height=1.6in]{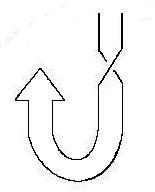}}
\caption{$ji=-k$}
\end{figure}

\noindent In this way the quaternions are realized by the behaviour of a belt attached to a wall in three dimensional space. See \cite{KP} for a more detailed discussion of the belt trick and its relationship with the quaternions.
\bigbreak

\section{The Octonions}

\begin{figure}[!h]
  \begin{center}
    \includegraphics[height=1.6in]{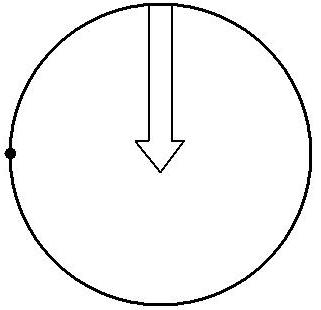}
  \end{center}
\caption{Identity} \label{identity}
\end{figure}

We construct our model for the octonions in a similar manner to the model for the quaternions.  Rather than using a belt, we will instead use a two toned ribbon (black on the back, and white on the front) with an arrowhead attached to one end (much as our belt had a buckle).  The other end is then attached to the interior of a ring (much as our belt was attached to a wall).  Lastly on the side of the ring we affix a flag that allows us to keep track of the orientation of the ring.  We will describe all operations with respect to viewing the ring from above the plane in which it lies with the white side of the ribbon facing us at the point where it is attached to the ring. Additionally, as in the model of the quaternions, we will perform all composite operations from left to right. We then begin with the identity as shown in figure \ref{identity}.

\begin{figure}[!h]
  \begin{center}
\subfloat[$i$]{\includegraphics[height=1.6in]{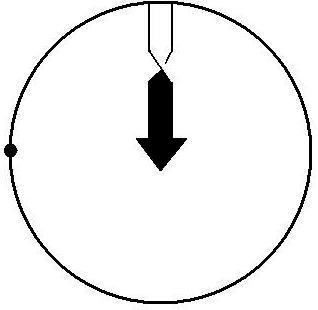}}\qquad    \subfloat[$j$]{\includegraphics[height=1.6in]{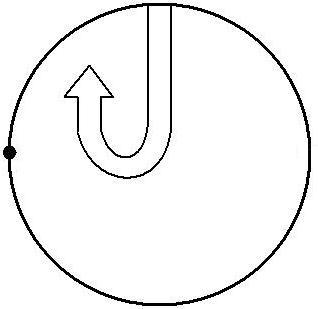}}\qquad
\subfloat[$k$]{\includegraphics[height=1.6in]{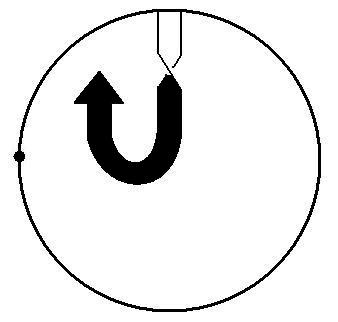}}
  \end{center}
\caption{$i$, $j$ and $k$} \label{quats}
\end{figure}

We can also then bring with us from our model of the quaternions $i$,$j$ and $k$ as shown in figure \ref{quats}.

Our operation $L$ is defined by  moving the flag to the opposite side of the ring and our operations $Li$, $Lj$ and $Lk$ are defined by  corresponding {\it rotations of the external hoop  while holding the ribbon stationary}, together with a reversal of the colour of the ribbon.  Note that rotating the external hoop will create twists in the ribbon. After performing each operation we perform global rotations to return our setup to standard form - the white side of the ribbon attached to the top of the hoop. No extra twists
are created by these global rotations.

\begin{figure}[!h]
  \begin{center}
\subfloat[$L$]{\includegraphics[height=1.6in]{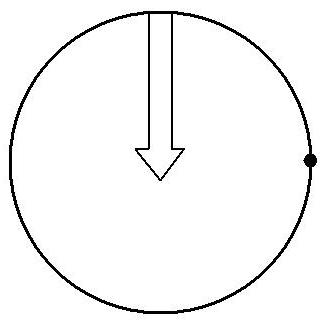}}\qquad \subfloat[$Li$]{\includegraphics[height=1.6in]{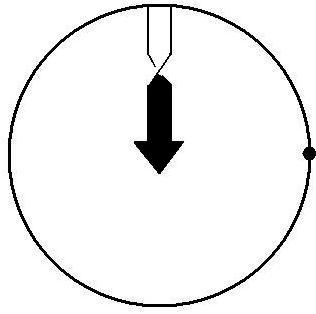}}\qquad     \subfloat[$Lj$]{\includegraphics[height=1.6in]{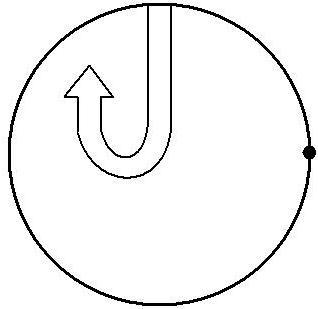}}\qquad
\subfloat[$Lk$]{\includegraphics[height=1.6in]{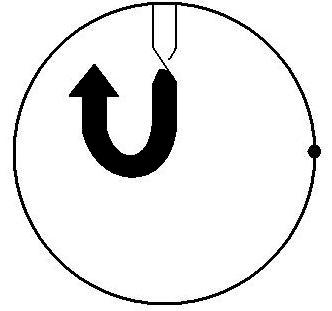}}
  \end{center}
\caption{$L$,$Li$, $Lj$ and $Lk$} \label{octs}
\end{figure}

We then have the form of $L$, $Li$, $Lj$ and $Lk$ applied to the identity given by figure \ref{octs}.

This gives us the basic form of the operations.  The issue is that some ambiguity exists in how to apply each operation given different configurations of the flag, colours and locations of the arrow.

\section{The Rules}
To introduce the consistent rules we will introduce some terminology.  We will refer to an arrowhead \textit{pointing up} if we have a configuration such as that for $j$ (see figure \ref{quats}) where the arrowhead points towards the top half of the hoop.  We will call a state \textit{flag-right} if when viewed in standard form the flag is on the right side of the hoop.  Lastly, we will say a state has a \textit{black arrowhead} if there is a twist in the ribbon in such a manner that the ribbon attaches to the arrowhead with the black side of the ribbon facing up.

\subsection{The Quaternion Rules}
\begin{enumerate}
\item The operation $i$ is a clockwise rotation of the arrowhead with respect to the hoop (through an axis through the arrowhead to the top of the hoop).  {\it This direction of rotation is reversed if the state is flag-right or if the arrowhead is pointing up, but not for both. }

\item The operation $j$ is a clockwise rotation of the arrowhead as viewed from above the hoop for any configuration of state.

\item The operation $k$ is performed by flipping the arrowhead under the ribbon, {\it unless the arrowhead is pointing up} - in which case we instead flip the arrowhead over the ribbon.

\end{enumerate}

It is important to note that excluding the reversal of $i$ for a state being flag-right that these are the standard rules for the model of the quaternions.
\bigbreak

\subsection{Rules for $L$ and the other Octonions}
\begin{figure}[!h]
  \begin{center}
    \subfloat[]{\includegraphics[height=1.3in]{identity.jpg}\label{lkid}}
    \subfloat[]{\includegraphics[height=1.3in]{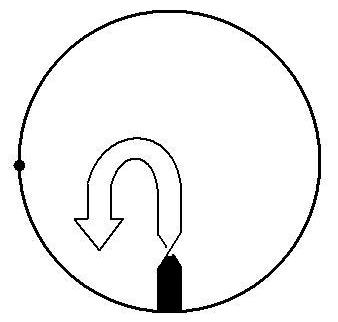}\label{lkrot}}
    \subfloat[]{\includegraphics[height=1.3in]{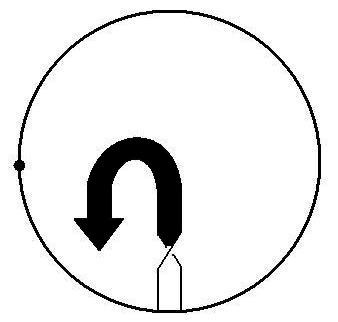}\label{lkcolour}}
    \subfloat[]{\includegraphics[height=1.3in]{Lk.jpg}\label{lk}}
  \end{center}
\caption{Step by step application of $Lk$}\label{lksteps}
\end{figure}

\begin{enumerate}

\item The operation $L$ is defined by  switching the side of the hoop that the flag is attached to, and {\it performing a full $2\pi$ rotation of the hoop (or - alternately - the arrowhead) if the arrowhead is pointing up or if the state is flag-right, but not for both.}

\item $Li$ takes the form - when looking through from the base of the hoop to the root of the ribbon - of rotating the hoop clockwise and reversing the colouring of the ribbon.  If the arrowhead is pointing up, we instead rotate counterclockwise.  Looking to figure \ref{listeps} we see this process at each step: from the identity we first rotate the hoop clockwise (figure \ref{lirot}), then we reverse the colour of the ribbon (figure \ref{licolour}) and then - as we are already now in standard form, with the hoop showing the attachment to the white ribbon - we are finished.

\begin{figure}[!h]
  \begin{center}
    \subfloat[]{\includegraphics[height=1.3in]{identity.jpg} \label{liid}}
    \subfloat[]{\includegraphics[height=1.3in]{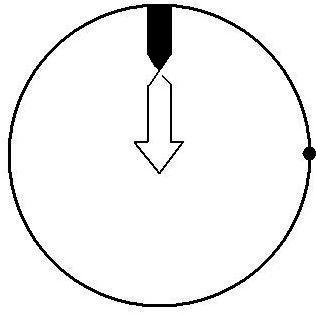}\label{lirot}}
    \subfloat[]{\includegraphics[height=1.3in]{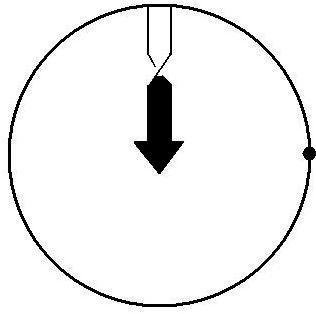}\label{licolour}}
  \end{center}
\caption{Step by step application of Li}\label{listeps}
\end{figure}

\item $Lj$ takes the form of rotating the hoop clockwise - from the standard view - and reversing the colouring of the ribbon.  If the state has a black arrowhead, or is flag-right we reverse the direction of rotation, but not if the state is both.  Taking $Lj$ step by step we get (in figure \ref{ljsteps}): a rotation of the hoop clockwise in the plane (figure \ref{ljrot}), reversing the colouring (figure \ref{ljcolour}) and then lastly performing a rotation of the total system to put it into standard form (figure \ref{lj}).

\begin{figure}[!h]
  \begin{center}
    \subfloat[]{\includegraphics[height=1.3in]{identity.jpg} \label{ljid}}
    \subfloat[]{\includegraphics[height=1.3in]{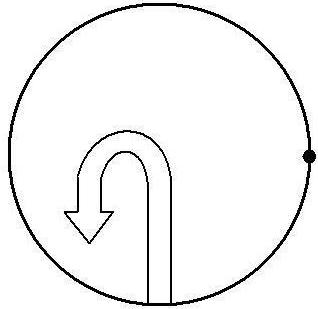}\label{ljrot}}
    \subfloat[]{\includegraphics[height=1.3in]{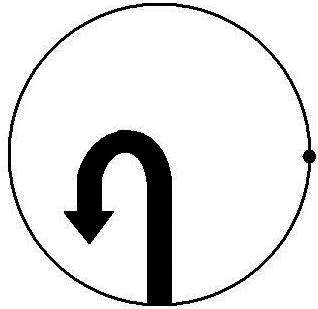}\label{ljcolour}}
    \subfloat[]{\includegraphics[height=1.3in]{Lj.jpg}\label{lj}}
  \end{center}
\caption{Step by step application of Lj}\label{ljsteps}
\end{figure}

\item $Lk$ takes the form of a clockwise rotation of the hoop - when viewing the hoop from the left - and reversing the colouring of the ribbon.  We reverse the direction of rotation for each of a black arrowhead, flag-right or an arrowhead pointing up (i.e. we would then have a clockwise rotation for a state with any two, and a counterclockwise rotation for a state that was all three).  The step by step process of applying $Lk$ to the identity is given by: rotating the hoop (figure \ref{lkrot}), changing the colour (figure \ref{lkcolour}), and then rotating the hoop and ribbon together to put it into standard form (figure \ref{lk}).

\end{enumerate}

\begin{figure}[!h]
  \begin{center}
  \subfloat[]{\includegraphics[height=1.6in]{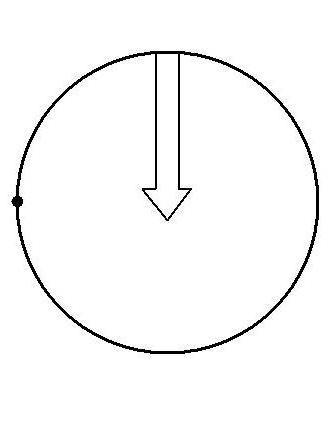}}\qquad
    \subfloat[]{\includegraphics[height=1.6in]{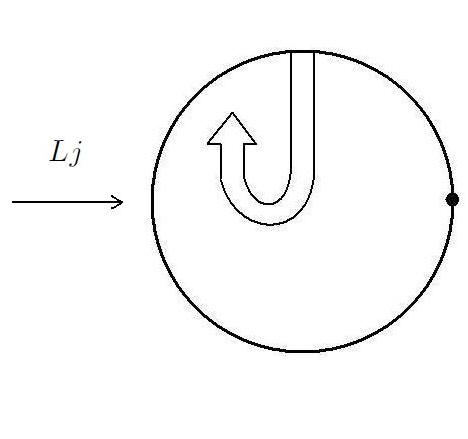}} \qquad
    \subfloat[]{\includegraphics[height=1.6in]{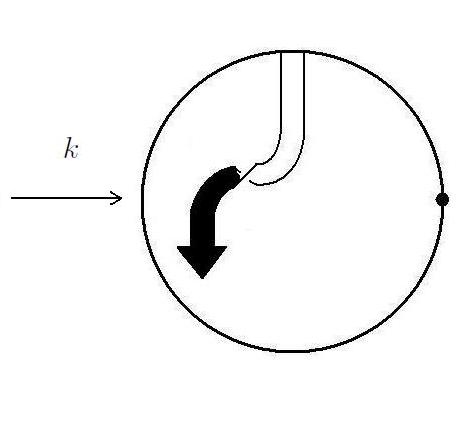}\label{jli1}}\qquad
    \subfloat[]{\includegraphics[height=1.6in]{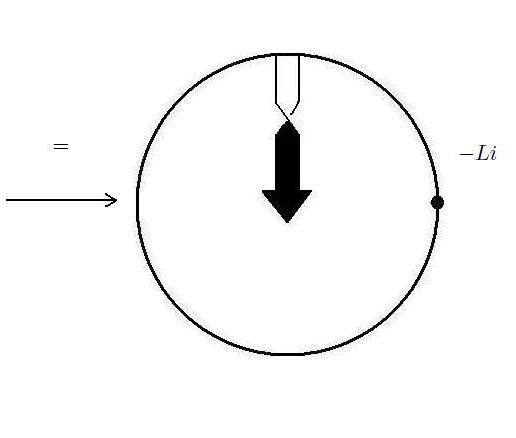}\label{jli2}}
  \end{center}
\caption{Calculation of $(Lj)k$}\label{Ljk}
\end{figure}

\begin{figure}[!h]
  \begin{center}
  \subfloat[]{\includegraphics[height=1.6in]{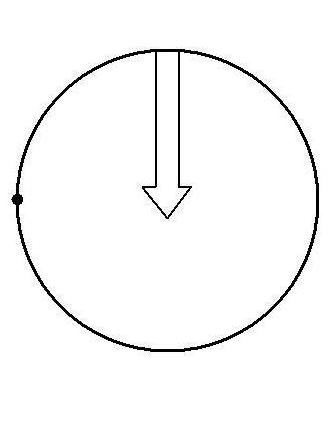}}\qquad
    \subfloat[]{\includegraphics[height=1.6in]{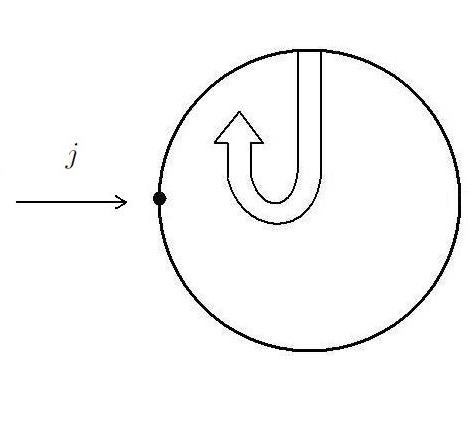}} \qquad
    \subfloat[]{\includegraphics[height=1.6in]{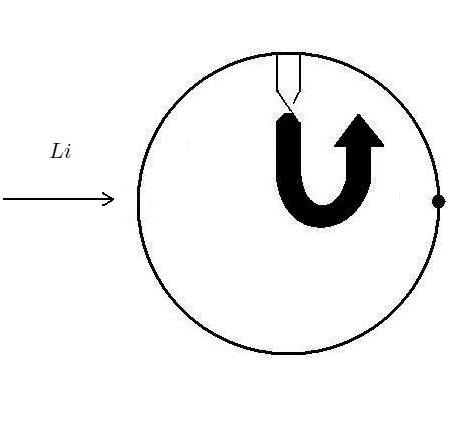}\label{jli1}}\qquad
    \subfloat[]{\includegraphics[height=1.6in]{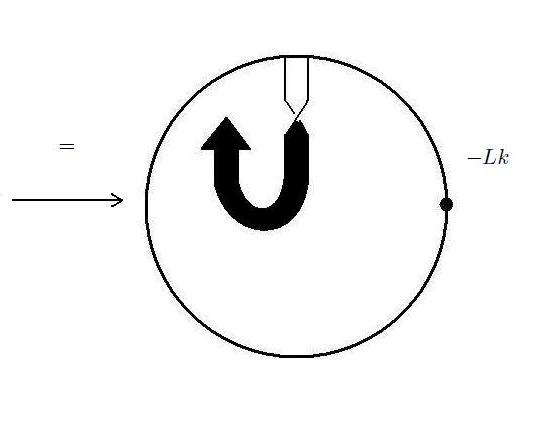}\label{jli2}}
  \end{center}
\caption{Calculation of $j(Li)$}\label{jli}
\end{figure}

To demonstrate these rules in practice (not just multiplied to the identity) and simultaneously to demonstrate that we have captured the non-associativity of the octonions, we'll perform two example calculations: $(Lj)k$ and $j(Li)$.  To perform the first (see figure \ref{Ljk}), we begin with $Lj$ as performed on the identity, and then we carry out $k$ - as $Lj$ has an upward pointing arrow, the direction of $k$'s rotation is reversed from its application on the identity, the result - after we remove deformation - is $(Lj)k = L(kj) = L(-i) = -Li$, comparing this to $L(jk) = Li$ and we find that we've captured the non-associativity in this scenario.  Next, we'll take $j(Li)$ (see figure \ref{jli}), here we begin by performing $j$, and then perform $Li$ which - as the arrowhead is pointing up - is a counterclockwise rotation of the ribbon, followed by a reversal of colouring (see figure \ref{jli1}).  We now have the arrowhead to the right of the ribbon, we can drag the ribbon to the other side of the arrowhead (this reverses the crossing) and we arrive at $-Lk$ which as $j(Li) = L(ji) = -L(k)$ is as desired.

\section{Conclusion}

We have now given an assignment of an operation to each of the octonions acting upon a state of our belt-hoop apparatus.  To demonstrate that this is indeed a model of the octonions, one only needs to check that any string of elements of the octonions can be equivalently resolved by the standard rules of octonionic multiplication or by applying them on our state in order - we stress here that the operations should be performed from left to right (i.e. $L i$ is $L$ followed by $i$).

This model of the octonions introduces several questions.  The original belt model of the quaternions is strongly related to the quaternions being a representation of SU(2), and SU(2) being a double cover of the rotation group SO(3).  The fact that this model of the octonions is an extension of the quaternionic model leads to the question of whether an analogue to the relationship with SU(2) and SO(3) exists.  In particular the form of $L$ - in some sense resembling a parity operation - plays to these speculations.

Extending on the relationship of the quaternions with SU(2) is the question of whether this model could provide illumination to attempts to use the octonions to construct the standard model of particle physics - such as the attempt in \cite{Furey:2010fm}.  Here again the resemblance of $L$ to parity inversion is suggestive of something more profound. We will continue these considerations in a sequel to the
present paper.

\section{Acknowledgements}
We would like to thank both Sundance Bilson-Thompson and Lee Smolin for their helpful discussions regarding this work.

Research at the Perimeter Institute for Theoretical Physics is supported in part by the Government of Canada through NSERC and by the Province of Ontario through MRI.

\end{document}